\documentclass[10pt]{article}
\usepackage{cite,makeidx,amscd,latexsym,graphicx,graphics,shapepar,hyperref,amsmath,amsthm,amsopn,amstext,amsfonts,amsbsy}
\newtheorem{thm}{Theorem}[section]
\newtheorem{defn}[thm]{Definition}
\makeindex \textwidth=470pt 
\numberwithin{equation}{section}

\begin{document}
\author{Alireza Khalili Gol$^a$$^,$$^b$\\
${^a}$\textit{Department of Physics,~University of Pune,~Pune}-
$\textit{411007}$, \textit{India}
\\$^b$\textit{Department of Physics,~Islamic Azad
University-Oromiyeh Branch,}\\\textit{PO Box 969,~Oromiyeh,~Iran}\\
E-mail addresses:~alireza@physics.unipune.ernet.in}

\title{\textbf{Fractional Poisson Bracket}}
\maketitle \large
\begin{abstract}
In the present paper fractional Hamilton-Jacobi  equation has been
derived for  dynamical systems involving Caputo
derivative.~Fractional Poisson-bracket is introduced.~Further
Hamilton's canonical equations are formulated and quantum wave
equation corresponds to the fractional Hamilton-Jacobi equation is
suggested.~Illustrative examples have been worked out to explain
the formalism.
\end{abstract}\small \textit{Keywords}:~Hamilton-Jacobi quantized
equation;~Riemann-Liouville derivative;~Caputo
derivative;~Canonical
 transformation;~Generating function;~fractional Poisson-bracket

\section{Introduction}

In seminal papers Riewe~\cite{a8,a9} has formulated Lagrangian and
Hamiltonian mechanics to include derivatives of fractional
order~\cite{a1,a2,a3,a4,a5,a6,a7}.~It has been shown that
Lagrangian involving fractional time derivatives leads to
equations of motion with non conservative classical forces such as
friction \cite{a8}. Motivated by this approach many researchers
have explored this area giving new insight into this
problem~\cite{a20,a21,a17,a10,a13,a11,a12}.~Agrawal \cite{a17} has
developed fractional calculus of variations dealing with problems
in which either the objective functional or the constraint
equations or both contain at least one fractional derivative
term.~Agrawal~\cite{a17} has dealt with Lagrangian involving
Riemann-Liouville (R-L) fractional derivatives.~R-L derivatives
are nonlocal.~R-L derivative of a constant is not zero,~and in
many applications it involves fractional initial conditions which
are nonphysical.~For these reasons Caputo derivative \cite{a6,a7}
has widely been used in recent literature.~Agrawal \cite{a10,a13}
in the recent papers has presented fractional Euler-Lagrange
equations involving Caputo derivatives.~In conclusion it is
emphasized that both (the R-L and Caputo) fractional derivatives
arise in the formulation,~even when the fractional variational
problem is defined only in terms of one type of derivative.~Thus
fractional boundary conditions may be necessary even when the
problem is defined in terms of Caputo derivative.~Further
fractional Hamiltonian formulation has been developed in terms of
Caputo derivatives by Baleanu and coworkers~\cite{a11,a12}.~As a
pursuit of this in the present paper we investigate the fractional
Hamiltonian involving Caputo derivative and derive the
Hamilton-Jacobi equations.~Poisson brackets constitute important
part of Hamiltonian mechanics.~Entire Hamiltonian mechanics can be
restated in terms of Poisson-bracket.~In view of this a
generalization of Poisson-bracket (fractional version) is
suggested.~Hamilton's canonical equations (fractional case) have
been expressed in terms of fractional Poisson bracket.~Further
fractional quantum wave equation is suggested.~Illustrative
examples are presented to explain the formalism.

\section{Fractional Calculus}
 Fractional calculus deals with generalizations of integer order derivatives integrals
 to arbitrary
 order.~In this section we present basic definitions and
 properties which will be used in the subsequent
 sections~\cite{a1,a2,a3,a4,a5}.
\begin{defn}~If~ $f(x)\in C[a,b]$~and
$\alpha>0$~then\cite{a2,a3,a4}
\begin{equation}
_{a}I^{\alpha}_{x}f(x):=\frac{1}{\Gamma(\alpha)}\int^{x}_{a}\frac{f(t)}
{(x-t)^{1-\alpha}}dt,~~~~~x>a,
\end{equation}
\begin{equation}
_{x}I^{\alpha}_{b}f(x):=\frac{1}{\Gamma(\alpha)}\int^{b}_{x}\frac{f(t)}
{(x-t)^{1-\alpha}}dt,~~~~~x<b,
\end{equation}
are called as the left sided and the right sided Riemann-Liouville
fractional integral of order~$ \alpha$,~respectively.
\end{defn}
\begin{defn}
~Let $n-1\leq\alpha<n$,~then
\begin{equation}
_{a}D^{\alpha}_{x}f(x):=\frac{1}{\Gamma(n-\alpha)}\left(\frac{d}
{dx}\right)^{n}\int^{x}_{a}\frac{f(t)}{(x-t)^{-n+\alpha+1}}dt,
\end{equation}
\begin{equation}
_{x}D^{\alpha}_{b}f(x):=\frac{1}{\Gamma(n-\alpha)}\left(-\frac{d}
{dx}\right)^{n}\int^{b}_{x}\frac{f(t)}{(t-x)^{-n+\alpha+1}}dt,
\end{equation}
are called as the left sided and the right sided Riemann-Liouville
fractional derivative of order~$\alpha$~respectively whenever the
RHS exists.
\end{defn}
\begin{defn}
~Let $f(x)\in C^{n}[a,b]$~and~$n-1\leq\alpha<n$,~then~\cite{a2,a4}
\begin{equation}
_{a}^{C}D^{\alpha}_{x}f(x)=_{a}I^{n-\alpha}_{x}D^{n}f(x)=\frac{1}
{\Gamma(n-\alpha)}\int^{x}_{a}(x-t)^{n-\alpha-1}\left(\frac{d}{dt}\right)^{n}f(t)dt,
~~~~~a<x<b
\end{equation}
\begin{equation}
_{x}^{C}D^{\alpha}_{b}f(x)=_{x}I^{n-\alpha}_{b}(-D)^{n}f(x)=\frac{1}
{\Gamma(n-\alpha)}\int^{b}_{x}(t-x)^{n-\alpha-1}\left(-\frac{d}{dt}\right)^{n}f(t)dt,
~~~~~a<x<b
\end{equation}
are called as the left sided and the right sided Caputo fractional
derivatives of order ~$\alpha$~respectively whenever the RHS
exists.
\end{defn}
Properties:~\cite{a13}\\
(i)
\begin{equation}
 _{a}^{C}D_{t}^{\alpha}(f(t)+g(t))=~_{a}^{C}D_{t}^{\alpha}f(t)+
 ~_{a}^{C}D_{t}^{\alpha}g(t).\label{eq:7}
\end{equation}
(ii)
\begin{equation}
_{a}^{C}D_{t}^{\alpha}c=0,~~~ c~~~ \textmd{is constant}.~~~~~~~~
\end{equation}
(iii)
\begin{equation}
 \int_{a}^{b}~[_{a}^{C}D_{t}^{\alpha}f(t)] g(t)dt=
 \int_{a}^{b}f(t)[~_{t}^{C}D_{b}^{\alpha}g(t)]dt.
\end{equation}
(iv)
\begin{equation}
_{a}D_{t}^{\alpha}(t-a)^{\beta}=
\frac{\Gamma(\beta+1)}{\Gamma(\beta+1-\alpha)}(t-a)^{\beta-\alpha}~~~(\beta>\alpha).
\end{equation}
(v)
\begin{equation}
_{a}I_{t}^{\alpha}(t-a)^{\beta}=
\frac{\Gamma(\beta+1)}{\Gamma(\beta+1+\alpha)}(t-a)^{\beta+\alpha}.
\end{equation}
\begin{equation}
_{a}I_{t}^{\alpha}~_{a}D_{t}^{\alpha}x(t)=
x(t)-\sum_{j=1}^{n}\frac{(_{a}D_{t}^{\alpha-j}x)(a)}{\Gamma(\alpha+1-j)}
(t-a)^{\alpha-j}.
\end{equation}
\begin{equation}
_{t}I_{b}^{\alpha}~_{t}D_{b}^{\alpha}x(t)=
x(t)-\sum_{j=1}^{n}\frac{(_{t}D_{b}^{\alpha-j}x)(b)}{\Gamma(\alpha+1-j)}
(b-t)^{\alpha-j}.\label{eq:9}
\end{equation}
\begin{equation}
_{a}I_{t}^{\alpha}~_{a}^{C}D_{t}^{\alpha}x(t)=
x(t)-\sum_{j=0}^{n-1}\frac{(D^{j}x)(a)}{\Gamma(j+1)}
(t-a)^{j}.\label{eq:10}
\end{equation}

\begin{equation}
_{t}I_{b}^{\alpha}~_{t}^{C}D_{b}^{\alpha}x(t)=x(t)-\sum_{j=0}^{n-1}\frac{((-D)^{j}x)(b)}
{\Gamma(j+1)}(b-t)^{j}.
\end{equation}
\section{Fractional Mechanics} Agrawal and coworkers
\cite{a10,a12} have presented Euler-Lagrange equations for
fractional variational problems defined in terms of R-L and Caputo
derivatives.~In the following section we state a theorem regarding
Lagrangian involving left and right Caputo derivatives,~which will
be used in further discussion.
\begin{thm}
~Let~$J[q]$ be a functional of the form
\begin{equation}
J[q]=\int_{a}^{b}L(t,q,~_{a}^{C}D_{t}^{\alpha}q,~_{t}^{C}D_{b}^{\beta}q)dt,\label{eq:11}
\end{equation}
where $0<\alpha,\beta<1$ and is defined on the set of functions
$f(x)$ which have continuous left Caputo fractional derivative
(LCFD) of order $\alpha$ and right Caputo fractional derivative
(RCFD) of order $\beta$ in $[a,b]$.~ A necessary condition for
$J[q]$ to have an extremum for a given function $q(t)$ is that
$q(t)$ satisfies the generalized Euler-Lagrange equation:
\begin{equation}
\frac{\partial L}{\partial q}+~_{t}D_{b}^{\alpha}\frac{\partial
L}{\partial~
_{a}^{C}D_{t}^{\alpha}q}+~_{a}D_{t}^{\beta}\frac{\partial
L}{\partial~_{t}^{C}D_{b}^{\beta}q}=0,~~~
t~\epsilon~[a,b],\label{eq:12}
\end{equation}
and the transversality conditions:
\begin{equation}
~\left[_{t}D_{b}^{\alpha-1}\frac{\partial L}{\partial~
_{a}^{C}D_{t}^{\alpha}q}-~_{a}D_{t}^{\beta-1}\frac{\partial
L}{\partial~_{t}^{C}D_{b}^{\beta}q}\right]\eta(t)|_{a}^{b}=0,\label{eq:13}
\end{equation}\end{thm}
where~$ _{t}D_{b}^{\alpha-1}$~denotes the fractional integral of
order $1-\alpha$.~See \cite{a10}~for a proof.

\section{Fractional Canonical Transformations and Generating
Functions } In this section we present the Hamiltonian formulation
involving Caputo fractional derivatives. Consider the fractional
Lagrangian  given in equation~(\ref{eq:11}). Then the canonical
momenta $p_{\alpha}$ and
$p_{\beta}$ are \\
\begin{equation}
p_{\alpha}=\frac{\partial
L}{\partial~_{a}^{C}D_{t}^{\alpha}q},~~p_{\beta}=\frac{\partial
L}{\partial~_{t}^{C}D_{b}^{\beta}q},\label{eq:14}
\end{equation}
where~$p_{\alpha}$ and $p_{\beta}$ are independent.~The fractional
canonical
Hamiltonian is \\
\begin{equation}
H=p_{\alpha}~_{a}^{C}D_{t}^{\alpha}q+~p_{\beta}~_{t}^{C}D_{b}^{\beta}q-L.\label{eq:15}
\end{equation}
Taking total differential of (\ref{eq:15}) and  using (\ref{eq:14}), we obtain \\
\begin{equation}
dH=dp_{\alpha}~_{a}^{C}D_{t}^{\alpha}q+~dp_{\beta}~_{t}^{C}D_{b}^{\beta}q-\frac{\partial
L}{\partial q}dq-\frac{\partial L}{\partial t}~dt.\label{eq:16}
\end{equation}
Taking into account the fractional Euler-Lagrange
equations~(\ref{eq:12}) we
get\\
\begin{equation}
dH=dp_{\alpha}~_{a}^{C}D_{t}^{\alpha}q+
~dp_{\beta}~_{t}^{C}D_{b}^{\beta}q+(~_{t}D_{b}^{\alpha}p_{\alpha}+
~_{a}D_{t}^{\beta}p_{\beta})dq-\frac{\partial L}{\partial
t}~dt.\label{eq:17}
\end{equation}
Equation (\ref{eq:17}) shows that $H$ is a function of $p_{\alpha},~p_{\beta},~q$ and $t$.~Comparing total differential of $H$ equation (\ref{eq:17}) we have:\cite{a12}\\
\begin{equation}
\frac{\partial H}{\partial t}=-\frac{\partial L}{\partial
t},~~\frac{\partial H}{\partial
p_{\alpha}}=~_{a}^{C}D_{t}^{\alpha}q,~~~\frac{\partial H}{\partial
p_{\beta}}=~_{t}^{C}D_{b}^{\beta}q,~~~\frac{\partial H}{\partial
q}=~_{a}D_{t}^{\beta}p_{\beta}+~_{t}D_{b}^{\alpha}p_{\alpha}.\label{eq:18}
\end{equation}
Transformation of ~$ q$, $p_{\alpha}$,~$p_{\beta}$ into new
variables~$Q(q,p_{\alpha},p_{\beta},t)$,~$P_{\alpha}(q,p_{\alpha},p_{\beta},t)$,
~$P_{\beta}(q,p_{\alpha},p_{\beta},t)$~is canonical if there
exists a new Hamiltonian
$\mathcal{H}(Q,P_{\alpha},P_{\beta},t)$~which satisfies modified
Hamilton  principle: \\
\begin{equation}
\delta\int_{a}^{b}(P_{\alpha}~_{a}^{C}D_{t}^{\alpha}Q~+
~P_{\beta}~_{t}^{C}D_{b}^{\beta}Q~-H)dt=0.
\end{equation}
As $q,~p_{\alpha},~\textmd{and}~p_{\beta} $ are canonically
conjugate,~we have
\begin{equation}
\delta\int_{a}^{b}(p_{\alpha}~_{a}^{C}D_{t}^{\alpha}q~+
~p_{\beta}~_{t}^{C}D_{b}^{\beta}q~-\mathcal{H})dt=0.
\end{equation}
For these equations to hold,~the integrands must differ by a total
time derivative of an arbitrary function $G$,~hence \\
\begin{equation}
(P_{\alpha}~_{a}^{C}D_{t}^{\alpha}Q~+~P_{\beta}~_{t}^{C}D_{b}^{\beta}Q~-H)dt-
(p_{\alpha}~_{a}^{C}D_{t}^{\alpha}q~+~p_{\beta}~_{t}^{C}D_{b}^{\beta}q~-\mathcal{H})dt=dG.
\end{equation}
Since $G$ is not varied at the end points,~we get
\begin{equation}
\delta\int_{a}^{b}\frac{dG}{dt}dt=\delta[G(b)-G(a)]=0.\label{eq:22}
\end{equation}
The function $G$,~which completely determines the transformation
is called as a generating function.~For mechanics involving
fractional derivatives,~we introduce variables
~$\bar{q}_{\alpha},~\bar{q}_{\beta},
~\bar{Q}_{\alpha},~\bar{Q}_{\beta}$ satisfying
\begin{equation}
\frac{d\bar{q}_{\alpha}}{dt}=~_{a}^{C}D_{t}^{\alpha}q,~\frac{d\bar{q}_{\beta}}{dt}=
~_{t}^{C}D_{b}^{\beta}q,~\frac{d\bar{Q}_{\alpha}}{dt}=
~_{a}^{C}D_{t}^{\alpha}Q,~\frac{d\bar{Q}_{\beta}}{dt}=~_{t}^{C}D_{b}^{\beta}Q.\label{eq:23}
\end{equation}
For integer order derivatives,~these new coordinates are the same
as the usual canonical coordinates.~However,~while dealing with
fractional derivatives,~the coordinates~
$\bar{q}_{\alpha},~\bar{q}_{\beta},~\bar{Q}_{\alpha},~\bar{Q}_{\beta}$~will
not be canonical,~so all canonical expressions must be written in
terms of the original
coordinates$~_{a}^{C}D_{t}^{\alpha}q,~_{t}^{C}D_{b}^{\beta}q,
~_{a}^{C}D_{t}^{\alpha}Q,~_{t}^{C}D_{b}^{\beta}Q$.
\section{Canonical Transformation of The First Kind}
 For a generating function
$G(\bar{q}_{\alpha},\bar{q}_{\beta},\bar{Q}_{\alpha},\bar{Q}_{\beta},t)$
the transformation is
\begin{equation}
(p_{\alpha}~_{a}^{C}D_{t}^{\alpha}q~+~p_{\beta}~_{t}^{C}D_{b}^{\beta}q~-H)dt-
(P_{\alpha}~_{a}^{C}D_{t}^{\alpha}Q~+~P_{\beta}~_{t}^{C}D_{b}^{\beta}Q~-\mathcal{H})dt=
dG(\bar{q}_{\alpha},\bar{q}_{\beta},\bar{Q_{\alpha}},\bar{Q}_{\beta},t).\label{eq:24}
\end{equation}
We have
\begin{equation}
dG=\frac{\partial G}{\partial
\bar{q}_{\alpha}}d\bar{q}_{\alpha}+\frac{\partial G}{\partial
\bar{q_{\beta}}}d\bar{q}_{\beta}+\frac{\partial G}{\partial
\bar{Q}_{\alpha}}d\bar{Q}_{\alpha}+\frac{\partial G}{\partial
\bar{Q}_{\beta}}d\bar{Q}_{\beta}+\frac{\partial G}{\partial t}dt.
\end{equation}
Eqs.~(\ref{eq:22}-\ref{eq:24}) yield
\begin{equation}
\frac{\partial
G}{\partial\bar{q}_{\alpha}}=p_{\alpha},\frac{\partial
G}{\partial\bar{q}_{\beta}}=p_{\beta},\frac{\partial
G}{\partial\bar{Q}_{\alpha}}=-P_{\alpha},\frac{\partial
G}{\partial\bar{Q}_{\beta}}=-P_{\beta},\frac{\partial G}{\partial
t}=\mathcal{H}-H.
\end{equation}
\section{Canonical Transformation of The Second Kind}
Let~$\mathcal{S} $~be a generating function dependent
on~$\bar{q}_{\alpha},~\bar{q}_{\beta},~P_{\alpha},~P_{\beta},~t$.~From
equations (\ref{eq:23}),~(\ref{eq:24}) we have
\begin{equation}
dG=p_{\alpha}d\bar{q}_{\alpha}-P_{\alpha}d\bar{Q}_{\alpha}+
p_{\beta}d\bar{q}_{\beta}-P_{\beta}d\bar{Q}_{\beta}+
(\mathcal{H}-H)dt
\end{equation}
\begin{equation}
=p_{\alpha}d\bar{q}_{\alpha}-d(P_{\alpha}\bar{Q}_{\alpha})+
\bar{Q}_{\alpha}dP_{\alpha}+p_{\beta}d\bar{q}_{\beta}-
d(P_{\beta}\bar{Q}_{\beta})+\bar{Q}_{\beta}dP_{\beta}+(\mathcal{H}-H)dt.
\end{equation}
It is easy to observe that
\begin{equation}
d(G+P_{\alpha}\bar{Q}_{\alpha}+
P_{\beta}\bar{Q}_{\beta})=p_{\alpha}d\bar{q}_{\alpha}+\bar{Q}_{\alpha}dP_{\alpha}+
p_{\beta}d\bar{q}_{\beta}+\bar{Q}_{\beta}dP_{\beta}+(\mathcal{H}-H)dt.
\end{equation}
Let
$\mathcal{S}=G+P_{\alpha}\bar{Q}_{\alpha}+P_{\beta}\bar{Q}_{\beta}$
then
\begin{equation}
d\mathcal{S}=p_{\alpha}~d\bar{q}_{\alpha}+\bar{Q}_{\alpha}dP_{\alpha}+
p_{\beta}d\bar{q}_{\beta}+\bar{Q}_{\beta}dP_{\beta}+(\mathcal{H}-H)dt.
\end{equation}
Since $\mathcal{S}$~is a function of
$\bar{q}_{\alpha},~\bar{q}_{\beta},~P_{\alpha},~P_{\beta},~t$ we
can write
\begin{equation}
\frac{\partial\mathcal{S}}{\partial\bar{q}_{\alpha}}=p_{\alpha},~
\frac{\partial\mathcal{S}}{\partial\bar{q}_{\beta}}=p_{\beta},~
\frac{\partial\mathcal{S}}{\partial
P_{\alpha}}=\bar{Q}_{\alpha},~\frac{\partial\mathcal{S}}{\partial
P_{\beta}}=\bar{Q}_{\beta},~\frac{\partial\mathcal{S}}{\partial
t}=\mathcal{H}-H.\label{eq:31}
\end{equation}
\section{Fractional Poisson Bracket} Hamiltonian mechanics can be
written in terms of Poisson brackets.~In the present section a
generalization of Poisson bracket has been introduced,~which is
useful for generalizing fractional mechanics involving Caputo
derivatives.
\begin{defn}\label{def:ooo}
If the functions $F(t,q,p_{\alpha},p_{\beta})$ and
$G(t,q,p_{\alpha},p_{\beta})$ depend on the position
coordinate,~fractional momenta and time,~fractional Poisson (FP)
bracket of $F$ and $G$,~denoted as $[F,G]_{FP}$ is defined to be:
\begin{equation}
[F,G]_{FP}=\frac{\partial F}{\partial q}(\frac{\partial
G}{\partial p_{\alpha}}+ \frac{\partial G}{\partial
p_{\beta}})-\frac{\partial G}{\partial q}(\frac{\partial
F}{\partial p_{\alpha}}+ \frac{\partial F}{\partial p_{\beta}}).
\end{equation}
\end{defn}
The following properties can be observed:\\
(a)
\[
[F,G]_{FP}=-[G,F]_{FP},
\]
(b)
\[
[F_{1}+F_{2},G]_{FP}=[F_{1},G]_{FP}+[F_{2},G]_{FP},
\]
(c)
\[
[F_{1},[F_{2},F_{3}]_{FP}]_{FP}+[F_{2},[F_{3},F_{1}]_{FP}]_{FP}+[F_{3},
[F_{1},F_{2}]_{FP}]_{FP}=0~~~~Jacobi's~ identity,
\]
(d)
\[
[F,q]_{FP}=-(\frac{\partial F}{\partial p_{\alpha}}+\frac{\partial
F}{\partial p_{\beta}}),
\]
(f)
\[
[F,p_{\alpha}]_{FP}=[F,p_{\beta}]_{FP}=\frac{\partial F}{\partial
q},
\]
(g)
\[
[q,q]_{FP}=[p_{\alpha},p_{\alpha}]_{FP}=[p_{\alpha},p_{\beta}]_{FP}=0,
\]
\[
[p_{\alpha},q]_{FP}=[p_{\beta},q]_{FP}=-1.
\]
\section{Hamilton's Canonical Equation In terms of  Poisson
Bracket}
 Hamilton's canonical equations in terms of Poisson
brackets can be expressed as follows
\begin{equation}
[q,H]_{FP}=(\frac{\partial H}{\partial p_{\alpha}}+\frac{\partial
H}{\partial p_{\beta}}),
\end{equation}
\begin{equation}\label{eq:lll}
[p_{\alpha},H]_{FP}=[p_{\beta},H]_{FP}=-(_{a}D^{\beta}_{t}p_{\beta}+
_{t}D^{\alpha}_{b}p_{\alpha})=-\frac{\partial H}{\partial q}.
\end{equation}

\section{Fractional Quantum Wave Equation } As in conventional
mechanics, the Hamilton-Jacobi (H-J) equation results from a
canonical transformation for which the new variables are constant.
For integer-order derivatives, such a transformation will follow
automatically if the new Hamiltonian $\mathcal{H}$ is identically
zero,~since from the equations of motion we then have
\begin{equation}
\dot{Q}=\frac{\partial \mathcal{H}}{\partial
P}=0,\dot{P}=-\frac{\partial\mathcal{H}}{\partial Q}.
\end{equation}
For fractional derivatives,~we can derive a similar relationship
by putting
\begin{equation}
\frac{\partial\mathcal{S}}{\partial
t}+H(q,p_{\alpha},p_{\beta},t)=0.\label{eq:33}
\end{equation}
In view of (\ref{eq:31}),~(\ref{eq:33}) yields fractional version
of H-J equation~\textit{i.e}
\begin{equation}
\frac{\partial\mathcal{S}}{\partial
t}+H(q,\frac{\partial\mathcal{S}}
{\partial\bar{q}_{\alpha}},\frac{\partial\mathcal{S}}
{\partial\bar{q}_{\beta}},t)=0.\label{eq:34}
\end{equation}
 Hence the quantum wave equation corresponding to Hamilton-Jacobi involving  fractional
Caputo derivative is suggested to be~\cite{a8}
\begin{equation}
[H(q,-i\hbar\frac{\partial}{\partial\bar{q}_{\alpha}},-i\hbar\frac{\partial}
{\partial\bar{q}_{\beta}},t)]\psi=i\hbar\frac{\partial
\psi}{\partial t},\label{eq:35}
\end{equation}
where $\psi $ is a wave function.
\section{Examples}
\textbf{Example~1.} The total energy of the fractional oscillator
is given as\cite{a15}
\begin{equation}
H=\frac{1}{2}kx^{2}+\frac{1}{2}m_{\alpha}(~_{a}^{C}D_{t}^{\alpha}x)^{2},~0<\alpha<1,~a<t<b,
\end{equation}
The generalized energy of  fractional oscillator in uniform
electric field $E$  will be
\begin{equation}
H_{FO}=q E
x+\frac{1}{2}kx^{2}+\frac{1}{2}m_{\alpha}(~_{a}^{C}D_{t}^{\alpha}x)^{2},~0<\alpha<1,~a<t<b,
\end{equation}
where $q$ the charge of oscillator,~ $m_{\alpha}=\gamma~m$ and
dimension of $\gamma$ is $T^{\alpha-1}$. In view of
Eq.~(\ref{eq:15}),~Lagrangian for FO is
\begin{equation}
L_{FO}=\frac{1}{2}m_{\alpha}(~_{a}^{C}D_{t}^{\alpha}x)^{2}-\frac{1}{2}kx^{2}-q
E x.
\end{equation}
The generalized Euler-Lagrange equation Eq.~(\ref{eq:12}) and the
transversality condition Eq.~(\ref{eq:13})~yield:
\begin{equation}
(a)~~~~~~~~
-qE-kx+m_{\alpha}~_{t}D_{b}^{\alpha}~_{a}^{C}D_{t}^{\alpha}x=0,~~~~~
~~~~~(b)~~~~~~~~~_{t}D_{b}^{\alpha-1}~_{a}^{C}D_{t}^{\alpha}x|_{t=b}=e_{1},
\label{eq:39}
\end{equation}
respectively. Eq.~(\ref{eq:39}) can be solved  to get explicit
expression for $x(t)$,~as follows:\\ Let
$\omega_{\alpha}^{2}=\frac{k}{m_{\alpha}}$
and $\frac{qE}{m_{\alpha}}=\gamma$.~The Euler-Lagrange equation then takes the form \\
\begin{equation}
_{t}D_{b}^{\alpha}~_{a}^{C}D_{t}^{\alpha}x=\omega_{\alpha}^{2}x+\gamma.~\label{eq:40}
\end{equation}
To find the solution of Eq.~(\ref{eq:40})~with initial condition
$x(0)=e_{0}$, we apply~$_{t}I_{b}^{\alpha}$ to both sides of
Eq.~(\ref{eq:40}).~Further using (\ref{eq:9})~and~
(\ref{eq:39})(b), we get
\begin{equation}\label{eq:41}
_{a}^{C}D_{t}^{\alpha}x(t)=\omega_{\alpha}^{2}~_{t}I_{b}^{\alpha}x(t)+
\frac{\gamma}{\Gamma(\alpha+1)}(b-t)^{\alpha}+\frac{e_{1}}{\Gamma(\alpha)}(b-t)^{\alpha-1}.
\end{equation}
Similarly, applying~ $_{a}I_{t}^{\alpha}$ to  both sides of
Eq.~(\ref{eq:41})~and using Eq.~(\ref{eq:10})~
 we obtain
\begin{equation}
x(t)=\omega_{\alpha}^{2}~_{a}I_{t}^{\alpha}~_{t}I_{b}^{\alpha}x+~_{a}I_{t}^{\alpha}
[\frac{e_{1}}{\Gamma(\alpha)}(b-t)^{\alpha-1}+\frac{\gamma}{\Gamma(\alpha+1)}(b-t)^{\alpha}]+
e_{0}.\label{eq:44}
\end{equation}as $(x(0)=e_{0})$.
Equation (\ref{eq:44}) can be thought of as Volterra-type integral
that has composite integral operators and can be solved to obtain
\begin{equation}
x(t)=(1-\omega_{\alpha}^{2}~_{a}I_{t}^{\alpha}~_{t}I_{b}^{\alpha})^{-1}(~_{a}I_{t}^{\alpha}
[\frac{e_{1}}{\Gamma(\alpha)}(b-t)^{\alpha-1}+\frac{\gamma}{\Gamma(\alpha+1)}(b-t)^{\alpha}]+
e_{0})
\end{equation}
\begin{equation}
=\sum_{j=0}^{\infty}(\omega_{\alpha}^{2}~_{a}I_{t}^{\alpha}~_{t}I_{b}^{\alpha})^
{j}(~_{a}I_{t}^{\alpha}
[\frac{e_{1}}{\Gamma(\alpha)}(b-t)^{\alpha-1}+\frac{\gamma}{\Gamma(\alpha+1)}(b-t)^{\alpha}]+
e_{0}).
\end{equation}
The generalized momentum for FO is
\begin{equation}
p_{\alpha}=\frac{\partial
L_{FO}}{\partial~_{a}^{C}D_{t}^{\alpha}q}=m_{\alpha}~_{a}^{C}D_{t}^{\alpha}x,
\end{equation}
and generalized Hamiltonian in terms of $p_{\alpha}$ is
\begin{equation}
H_{FO}=\frac{1}{2m_{\alpha}}p_{\alpha}^{2}+\frac{1}{2}kx^{2}+q E
x.\label{eq:48}
\end{equation}
Hence the Hamilton's equations (Eq.~(\ref{eq:18})) become
\begin{equation}
\frac{\partial H_{FO}}{\partial
p_{\alpha}}=\frac{p_{\alpha}}{m_{\alpha}}=~_{a}^{C}D_{t}^{\alpha}q,~~~~~\frac{\partial
H_{FO}}{\partial
q}=kx+qE=~_{t}D_{b}^{\alpha}~p_{\alpha},\label{eq:49}
\end{equation}
which are equivalent to generalized Euler-Lagrange equations given in Eq.~(\ref{eq:39})(a).\\
Further we present Hamilton-Jacobi equation for this
system.~Eq.~(\ref{eq:34})~yields the Hamilton-Jacobi Equation.
\begin{equation}
\frac{\partial \mathcal{S}_{FO}}{\partial
t}+\frac{1}{2m_{\alpha}}(\frac{\partial
\mathcal{S}_{FO}}{\partial\bar{x}_{\alpha}})^{2}+\frac{1}{2}kx^{2}+q
E x=0,\label{eq:50}
\end{equation}
where $\frac{d\bar{x}_{\alpha}}{dt}=~_{a}^{C}D_{t}^{\alpha}x(t)$.
Assume that a solution to Eq.~(\ref{eq:50}) is the form of
$\mathcal{S}_{FO}=S_{1}(t)+S_{2}( \bar{x}_{\alpha})$.~Then
\begin{equation}
\frac{1}{2m_{\alpha}}(\frac{dS_{2}}{d\bar{x}_{\alpha}})^{2}+\frac{1}{2}kx^{2}+q
E x=-\frac{dS_{1}}{dt}= \beta,
\end{equation}
where $\beta$ is a constant.~Therefore
\begin{equation}
\frac{dS_{2}}{d\bar{x}_{\alpha}}=\sqrt{2m_{\alpha}(\beta-\frac{1}{2}kx^{2}-q
E x)},S_{1}=-\beta t.
\end{equation}
So that
\begin{equation}
\mathcal{S}_{FO}=\bar{x}_{\alpha}\sqrt{2m_{\alpha}(\beta-\frac{1}{2}kx^{2}-q
E x)}-\beta t.
\end{equation}
Further we identify $\beta$ with the new momentum coordinate
$P_{\alpha}$.~Hence the new position variable $Q_{\alpha}$ will be
\begin{equation}
Q_{\alpha}=\frac{\partial\mathcal{S}_{FO}}{\partial
\beta}=\frac{\partial}{\partial \beta}
[\bar{x}_{\alpha}\sqrt{2m_{\alpha}(\beta-\frac{1}{2}kx^{2}-q E
x)}-\beta t]
\end{equation}
\begin{equation}
=\frac{m_{\alpha}}{\sqrt{2m_{\alpha}(H-\frac{1}{2}kx^{2}-q E
x)}}~\bar{x}_{\alpha}-t=\gamma,\label{eq:55}
\end{equation}
where $\gamma$ is a constant.~Solving Eq.~(\ref{eq:55}) for
$\bar{x_{\alpha}}$~we  get
\begin{equation}
m_{\alpha}
\bar{x}_{\alpha}=(\gamma+t)\sqrt{2m_{\alpha}(\beta-\frac{1}{2}kx^{2}-q
E x)}\label{eq:56}
\end{equation}
Further differentiating (\ref{eq:56})~with  respect to ~$t$,~we
get
\begin{equation}
m_{\alpha}\frac{d\bar{x}_{\alpha}}{dt}=\sqrt{2m_{\alpha}(\beta-\frac{1}{2}kx^{2}-q
E x)}=p_{\alpha}.\label{eq:57}
\end{equation}
In view of Eq.~(\ref{eq:34}),~we can identify $\beta$ with the
Hamiltonian $H$.~Then in
 view of (\ref{eq:48})
$m\frac{d\bar{x}_{\alpha}}{dt}=\sqrt{2m(\beta- \frac{1}{2}kx^{2}-q
E x)}=p_{\alpha}$.~But
\begin{equation}
m_{\alpha}~_{a}^{C}D_{t}^{\alpha}x=p_{\alpha}.\label{eq:58}
\end{equation}
Applying ~$_{t}D_{b}^{\alpha}$~to both sides of Eq.~(\ref{eq:58})
we get
\begin{equation}
m_{\alpha}~_{t}D_{b}^{\alpha}~_{a}^{C}D_{t}^{\alpha}x=~_{t}D_{b}^{\alpha}p_{\alpha}.
\end{equation}
Using Eq.~(\ref{eq:49}) we obtain
\begin{equation}
m_{\alpha}~_{t}D_{b}^{\alpha}~_{a}^{C}D_{t}^{\alpha}x=kx+qE,
\end{equation}
which is the same  equation that we derive from Euler-Lagrange and Hamilton equations.\\
This result can also be obtained using fractional Poisson
brackets.
\[
[x,H_{FO}]_{FP}=[x,\frac{1}{2}kx^{2}+q E
x+\frac{1}{2}m_{\alpha}(_{a}^{C}D_{t}^{\alpha}x)^{2}]_{FP}.
\]
Since the generalized momentum for FO is
\begin{equation}
p_{\alpha}=\frac{\partial
L_{FO}}{\partial~_{a}^{C}D_{t}^{\alpha}q}=m_{\alpha}~_{a}^{C}D_{t}^{\alpha}x,
\end{equation}
 we can write,
\begin{equation}
[x,H_{FO}]_{FP}=[x,\frac{1}{2}kx^{2}+q E x+
\frac{p_{\alpha}^{2}}{2m_{\alpha}}]_{FP}=\frac{p_{\alpha}}{m_{\alpha}}
=~_{a}^{C}D_{t}^{\alpha}x,
\end{equation}
and
\begin{equation}
[p_{\alpha},H_{FO}]_{FP}=[p_{\alpha},\frac{1}{2}kx^{2}+q E x+
\frac{p_{\alpha}^{2}}{2m_{\alpha}}]_{FP}=
-kx-qE=-~_{t}D_{b}^{\alpha}p_{\alpha}.
\end{equation}
These equations are the same as the Hamilton's equations
Eq.~(\ref{eq:49}).~Therefore,~both the methods yield the following
equation,~for fractional oscillator in uniform electric field
\begin{equation}
-kx-qE+m_{\alpha}~_{t}D_{b}^{\alpha}~_{a}^{C}D_{t}^{\alpha}x=0.
\end{equation}
Eq.~(\ref{eq:35})~yields the following wave equation corresponding
to fractional oscillator in the uniform electric field $E$
\begin{equation}
(\frac{1}{2m_{\alpha}}(-i\hbar\frac{\partial}{\partial\bar{x}_{\alpha}})^{2}+
\frac{1}{2}kx^{2}+q E x)\psi=i\hbar\frac{\partial \psi}{\partial
t}.\label{eq:61}
 \end{equation}
The wave function for the quantized FO is of the form
\begin{equation}
\psi(x,\bar{x}_{\alpha},t)=
A(x,\bar{x}_{\alpha},t)~e^{\frac{i}{\hbar}\mathcal{S}_{FO}
(x,\bar{x}_{\alpha},t)},\label{eq:62}
\end{equation}
where $A(x,\bar{x}_{\alpha},t)$ and
~$\mathcal{S}_{FO}(x,\bar{x}_{\alpha},t)$ are
 amplitude and phase respectively.
Substituting Eq.~(\ref{eq:62})~in~Eq.~(\ref{eq:61}) we get
\begin{equation}
-\frac{\hbar^{2}}{2m_{\alpha}}(\frac{\partial
A}{\partial\bar{x}_{\alpha}})^{2}+
\frac{1}{2m_{\alpha}}A(\frac{\partial\mathcal{S}_{FO}}{\partial\bar{x}_{\alpha}})^{2}
+(\frac{1}{2}kx^{2}+q E x)A=i\hbar\frac{\partial A}{\partial t}-
A\frac{\partial \mathcal{S}_{FO}}{\partial t}.
\end{equation}
Separating  this expression into real and imaginary parts,~we get
the two equations:
\begin{equation}
\frac{1}{2m_{\alpha}}(\frac{\partial\mathcal{S}_{FO}}{\partial\bar{x}_{\alpha}})^{2}+
\frac{\partial \mathcal{S}_{FO}}{\partial t}+\frac{1}{2}kx^{2}+q E
x=\frac{\hbar^{2}}{2m_{\alpha}} (\frac{\partial
A}{\partial\bar{x}_{\alpha}})^{2}
~~~~~\textmd{and}~~~\frac{\partial A}{\partial t}=0.\label{eq:65}
\end{equation}
 Eq.~(\ref{eq:65}) reduces to fractional Hamilton-Jacobi,~when
 $\hbar=0$.\\
 \\
\textbf{Example~2.}~Consider FO with dissipative force:
\begin{equation}
F(x)=-\gamma~ _{t}^{C}D_{b}^{\beta}x
\end{equation}
where dimension of $\gamma$ is~$MLT^{-(1+\beta)}$.~Using potential
corresponding to this force we have Lagrangian as~\cite{a8}:
\begin{equation}
L=\frac{1}{2}m_{\alpha}(~_{a}^{C}D_{t}^{\alpha}x)^{2}-\frac{1}{2}kx^{2}-
i\frac{\gamma}{2(-1)^\frac{\beta}{2}}(~_{t}^{C}D_{b}^{\frac{\beta}{2}}x)^{2}.
\end{equation}
In  view of  Eq.~(\ref{eq:12}) we obtain generalized
Euler-Lagrange equation as
\begin{equation}\label{o78}
-kx+m_{\alpha}~_{t}D_{b}^{\alpha}~_{a}^{C}D_{t}^{\alpha}x-
i\frac{\gamma}{(-1)^\frac{\beta}{2}}~_{a}D_{t}^\frac{\beta}{2}~_{t}^{C}D_{b}^\frac{\beta}{2}x=0.
\end{equation}
The generalized momenta are
\begin{equation}\label{mmmm}
p_{\alpha}=m_{\alpha}~_{a}^{C}D_{t}^{\alpha}x~~~~
\textmd{and}~~~~~~p_\frac{\beta}{2}=-i\frac{\gamma}{(-1)^\frac{\beta}{2}}~_{t}^{C}D_{b}^\frac{\beta}{2}x.
\end{equation}
The corresponding Hamiltonian is
\begin{equation}
H=\frac{1}{2}m_{\alpha}(_{a}^{C}D_{t}^{\alpha}x)^{2}+\frac{1}{2}kx^{2}+
i\frac{\gamma}{2(-1)^\frac{\beta}{2}}~(_{t}^{C}D_{b}^\frac{\beta}{2}x)^{2}.
\end{equation}
Using Eqs.~(\ref{mmmm}) the Hamiltonian for this system takes the
form
\begin{equation}
H=\frac{p_{\alpha}^{2}}{2m_{\alpha}}+
\frac{1}{2}kx^{2}+\frac{p_\frac{\beta}{2}^{2}}{2i\gamma(-1)^\frac{\beta}{2}}~.
\end{equation}
Using Hamilton's canonical  Poisson bracket we have
\begin{equation}\label{imnh}
[p_{\alpha},H]=[p_{\alpha},\frac{p_{\alpha}^{2}}{2m_{\alpha}}+
\frac{1}{2}kx^{2}+\frac{p_\frac{\beta}{2}^{2}}{2i\gamma(-1)^\frac{-\beta}{2}}]=-kx=
[p_\frac{\beta}{2},H]=
-(_{a}D^\frac{\beta}{2}_{t}p_\frac{\beta}{2}+
_{t}D^{\alpha}_{b}p_{\alpha}).
\end{equation}
Substituting Eqs.~(\ref{mmmm}) in Eq.~(\ref{imnh})~we arrive at
Eq.~(\ref{o78}).
\section{Results and Conclusions}
Fractional mechanics describes both conservative and
non-conservative systems.~With this motivation a  generalization
of Poisson bracket is introduced.~Further generalized Hamilton's
canonical equations have been derived in the case of Lagrangian
involving Caputo derivatives.~Subsequently Hamilton-Jacobi
equation has been derived and fractional quantum wave is
suggested.
\section{Acknowledgements}
The author is grateful to Professors V.~Daftardar-Gejji. and
~A.~D.~Gangal for useful discussions.

{}

\end{document}